\documentclass[aip,apl,reprint,amssymb,amsmath,superscriptaddress]{revtex4-1}
\usepackage{graphicx}
\usepackage{dcolumn}
\usepackage{bm}
\usepackage{textcomp}

\begin{document}

\title{Few-hole double quantum dot in an undoped GaAs/AlGaAs heterostructure}

\author{L. A. Tracy}
\email{latracy@sandia.gov}
\affiliation{Sandia National Laboratories, Albuquerque, New Mexico 87185, USA}

\author{T. W. Hargett}

\author{J. L. Reno}
\affiliation{Center for Integrated Nanotechnologies, Sandia National Laboratories, Albuquerque, New Mexico 87185, USA}

\date{\today}

\begin{abstract}
We demonstrate a hole double quantum dot in an undoped GaAs/AlGaAs heterostructure.
The interdot coupling can be tuned over a wide range, from formation of a large single dot to two well-isolated quantum dots.
Using charge sensing, we show the ability to completely empty the dot of holes 
and control the charge occupation in the few-hole regime.  
The device should allow for control of individual hole spins in single and double quantum dots in GaAs.
\end{abstract}

\pacs{}

\maketitle 

One of the leading candidates for a solid-state quantum bit is the spin of a single electron confined in a semiconductor \cite{Loss}.
The pioneering initial experiments demonstrating coherent control of individual electron spins in quantum dots utilized
high-mobility two-dimensional (2D) electron systems in GaAs/AlGaAs heterostructures \cite{Petta,Koppens}.  
The major source of decoherence in such experiments is coupling between electron spins and nuclear spins in the host GaAs semiconductor \cite{Petta,HansonRMP}.  
It has been proposed that hole spins in GaAs would be better suited for such experiments due to a lesser coupling between hole and nuclear spins 
\cite{LossDecoherence,Brunner,Yamamoto,Gammon}.  
The stronger spin-orbit interaction for holes, as compared to electrons, may also provide a means for electrical spin manipulation \cite{EDSR,ESDRInSbNanowire}.

To date, experiments on single spins in semiconductor quantum dots have primarily focused on electron spins.
Confinement of single hole spins in nanowire devices \cite{SiGeWires,ESDRInSbNanowire} 
and self-assembled quantum dots \cite{Brunner,Yamamoto,Gammon} has been demonstrated, 
but fabrication of quatum dots using conventional 2D heterostructures has potential advantages in terms of flexibility 
of tuning the confinement potential and for fabrication of mulitple dot devices \cite{conditional,Sachrajda}.
One of the main reasons for the lack of experiments on hole quantum dots in GaAs
is the difficulty of fabricating electrically stable nanostructures (such as quantum dots) in p-doped GaAs/AlGaAs heterostructures \cite{EnsslinSET,Grbic,Burke}.
However, undoped, enhancement-mode devices provide an alternate route to fabricating p-type nanostructures in GaAs.
High-mobility two-dimensional hole systems have already been demonstrated in undoped devices \cite{Kane,Lilly,TML} and
recent results have shown that a stable many-hole quantum dot can be formed in an enhancement-mode device in a (311)A oriented
heterostructure with a p-doped cap layer \cite{HamiltonSET}.

In this article, we report fabrication and measurement of a hole double quantum dot (DQD) in an undoped (100) oriented GaAs/AlGaAs heterostructure.
The mean free path in similarly processed bulk 2D devices at $T = 4$ K, at density of $p = 2 \times 10^{11}$ cm$^{-2}$ is $\sim$1 \textmu{}m, 
which is larger than typical nanostructure dimensions, aiding in the formation of low-disorder few-hole nanostructures.  
The gate design \cite{Ciorga} provides a high degree of tunability, 
allowing for independent control over individual dot occupation and tunnel barriers, 
as well as the ability to use a nearby quantum point contacts (QPCs) to sense dot charge occupation.
We show the ability to control the coupling between dots, tuning the device across the transition from one large dot to two well-isolated quantum dots. 
Using charge sensing, we determine the charge occupation of the DQD and demonstrate operation of the device in the few-hole regime.

\begin{figure}
\includegraphics[width = 88mm]{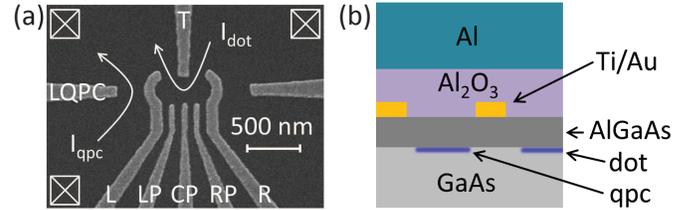}
\caption{\label{FIG. 1}  (a) Scanning electron micrograph of partially processed device, showing Ti/Au gates on GaAs/AlGaAs heterostructure surface used to form a quantum dot and
QPC charge sensor.  
(b) Sketch of cross section of left half of device.}
\end{figure}

Figure 1(a) shows a scanning electron micrograph of Ti/Au gates on the surface of a GaAs/AlGaAs heterostructure (VA0582) for a partially processed device.
Figure 1(b) shows a schematic cross section of the left half of the final device.
The upper Al gate is used to accumulate holes at the GaAs/Al$_{x}$Ga$_{1-x}$As (x = 0.5) interface, 100 nm below the heterostructure surface, as sketched in Fig.~1(b).
The lower, patterened Ti/Au (10 nm Ti, 40 nm Au) gates are used to locally deplete to define the QPC and DQD.
Ohmic contacts to the hole layer are formed via AuBe evaporation and anneal.
A 110 nm thick layer of Al$_2$O$_3$ grown via atomic layer deposition electrically isolates the upper gate from AuBe Ohmic contacts and lower Ti/Au gates \cite{TML,Willett}.
The device conductance is experimentally determined via standard low-frequency lock-in measurements with an rms ac source-drain bias of 50 - 100 \textmu{}V.  
For all measurements shown, the Al upper gate voltage is held constant at -6 V.
The device was measured in a $^3$He refrigerator with a temperature of $T = 380$ mK.

\begin{figure}
\includegraphics[width = 70mm]{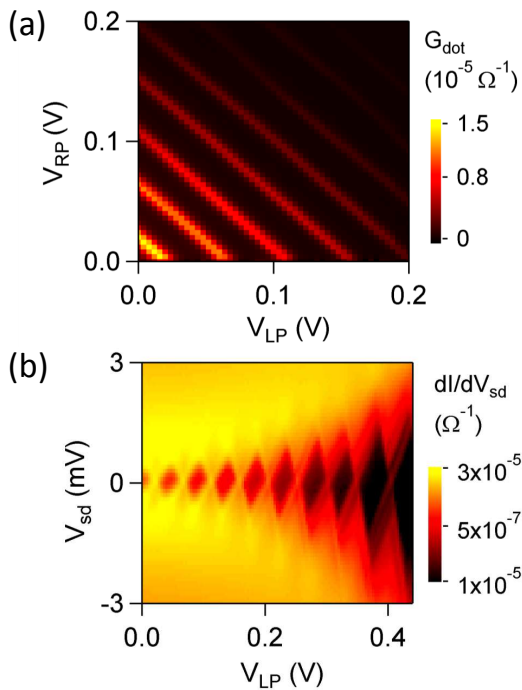}
\caption{\label{FIG. 2}  Dot conductance for a large single dot for fixed gate voltages 
$V_{CP}$ = 0 V, $V_{LQPC}$ = 0.3 V, $V_{L}$ = 0.2 V, and $V_{CP}$ = 0.275 V.
(a) Dot conductance $G_{dot}$ vs. $V_{LP}$ and $V_{RP}$ 
(b) $dI_{dot}/dV_{sd}$ versus $V_{sd}$ and $V_{LP}$ showing Coulomb diamonds.}
\end{figure}

Figure 2(a) shows quantum dot conductance versus left and right plunger gate voltages $V_{LP}$ and $V_{RP}$ with $V_{CP}$ = 0 V.
In this regime, the device behaves like a large single dot, with roughly equal capacitance between the dot and gates LP and RP, 
where $C_{dot-LP} = 3.5$ aF and $C_{dot-RP} = 3.6$ aF.
The data of Fig.~2(a) show no evidence of hysteresis or electrical instability, 
in contrast to mesoscopic devices fabricated in p-doped GaAs/AlGaAs heterostructures \cite{EnsslinSET,Grbic,Burke}.
In Fig.~2(b) we show a stability diagram with Coulomb diamonds for this dot.  The last visible diamond indicates a dot charging energy of $\sim1.6$ meV, 
and shows excited states with a spacing of roughly $E_{orb}\sim0.5$ meV.  
Using $E_{orb} \sim \pi \hbar^2/m^{\ast}l^2$, we obtain a rough estimate of the dot size \cite{KouwenhovenReview} $l \sim$ 100 nm.
Although the effective hole mass will depend on the details of the dot confinement potential \cite{Winkler}, 
here we use the effective mass for heavy holes in 2D systems in GaAs/AlGaAs single-interface heterostructures, $m^{\ast}_{HH} \sim 0.5m_{e}$ \cite{TMLcyc}.
For future devices, it would be of interest to utilize heterostructures with shallow 2D hole layers ($<100$ nm depth) \cite{shallowQD}.
This should help to decrease the size of the electrostatic confinement potential, 
which may be required in order to achieve similar orbital level spacings to those obtained in electron quantum dots,
since the heavy hole effective mass is larger than the electron effective mass in GaAs ($m^{\ast}_{HH} \sim 0.2m_e - 0.5m_e > m^{\ast}_{e} \approx 0.07m_e$ ).

\begin{figure}
\includegraphics[width = 85mm]{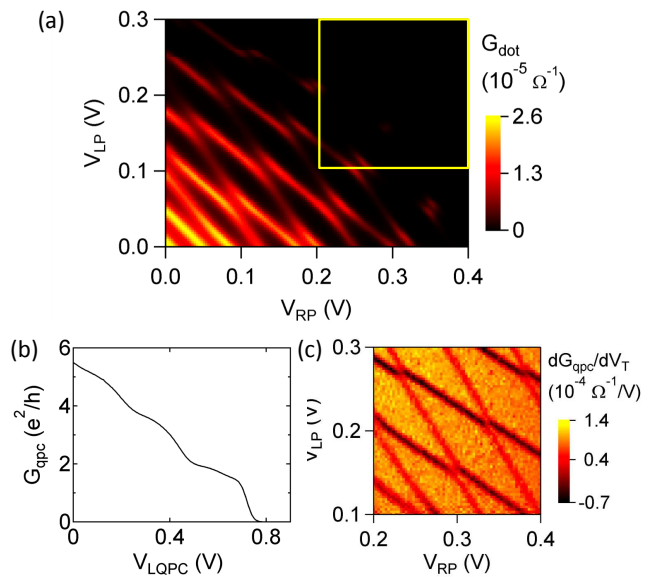} 
\caption{\label{FIG. 3} (a) Dot conductance $G_{dot}$ vs. $V_{LP}$ and $V_{RP}$ 
for fixed gate voltages $V_{CP}$ = 0.2 V, $V_{LQPC} = V_{L}$ = 0 V, and $V_{CP}$ = 0.5 V.  
The yellow box outlines the gate voltage region spanned in (c).
(b) Left QPC conductance vs. $V_{LQPC}$ for $V_{L}$ = 0 V.
(c) QPC transconductance $dG_{qpc}/dV_T$ vs. $V_{LP}$ and $V_{LP}$ with $V_{CP}$ = 0.2 V.  
$V_{LQPC}$ is varied from 0.43 to 0.4 V in order to maintain constant sensitivity.}
\end{figure}

Figure 3(a) shows quantum dot conductance versus $V_{LP}$ and $V_{RP}$ after increasing the center plunger voltage to $V_{CP}$ = 0.2 V.
Transport gradually evolves from single dot to double dot-like as the confinement is increased.
At the upper right corner of Fig.~3(a), the dot tunnel barriers become too opaque to measure conduction directly through the dot.
Figure 3(b) shows conductance through the left QPC versus $V_{LQPC}$.  
As expected for a QPC, the data show plateaux in the conductance, with the second-to-last plateau occuring near $2e^2/h$.
The last plateau occuring below $2e^2/h$ may be the so-called "0.7 structure",
which has been previously observed in electron \cite{Pepper} and hole QPCs \cite{Hamiltonp7,Ensslinp7}.
The precise conductance values, especially for the higher conductance plateaux,
 are likely affected by lead resistance since we use a two-terminal measurement.
In Fig.~3(c) we show the QPC transconductance $dG_{qpc}/dV_{T}$ versus $V_{LP}$ and $V_{RP}$ 
in the region of gate voltage indicated by the yellow box in Fig.~3(a).
In order to use the QPC to charge sense the occupation of the dot, 
we tune $V_{LQPC}$ in order to sit on the steep portion of the $G_{qpc}$ versus $V_{LQPC}$ curve below the last conductance plateau.
The data clearly show sensing of both left and right dot charge occupation, 
where single hole changes in the left dot occupation produce a larger change in $dG_{qpc}/dV_{T}$ than for the right dot 
due to the closer physical proximity between the left dot and QPC. 

\begin{figure*}
\includegraphics[width = 180mm]{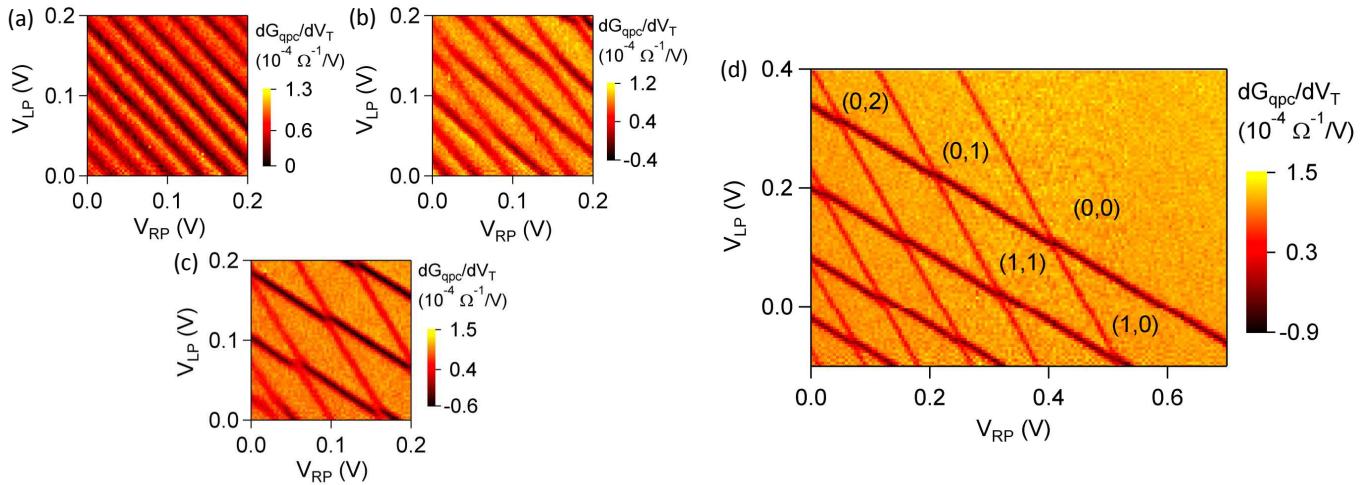}
\caption{\label{FIG. 4} Left QPC transconductance $dG_{qpc}/dV_T$ vs. $V_{LP}$ and $V_{LP}$ for (a) $V_{CP}$ = 0 V, 
(b) $V_{CP}$ = 0.2 V, (c) $V_{CP}$ = 0.4 V, and (d) $V_{CP}$ = 0.7 V.}
\end{figure*}

In Fig.~4(a)-(c) we show QPC transconductance versus left and right plunger gate voltages $V_{LP}$ and $V_{RP}$
at three different center plunger gate voltages $V_{CP} = 0$, 0.2, and 0.4 V.  The data
demonstrate the ability to use the CP gate to tune the DQD from a highly-coupled regime, 
where the transport is reminiscent of that expected for a large, single
dot, to a weakly-coupled regime where the charge sensing signal shows two well-isolated dots. 

Figure 4(d) shows a continuation of the charge sensing data to larger $V_{LP}$ and $V_{LP}$ voltages, for $V_{CP}$ = 0.7 V.
The absence of transitions in the charge sensing signal in the upper right region of the plot, for both the left and right dot, over a wide voltage range,
indicate that the DQD is empty.  
This allows us to label the various regions between charge transitions with DQD hole occupation ($N$,$M$), as shown in Fig.~4(d),
where $N$ ($M$) indicates the number of holes in the left (right) dot, respectively.

In conclusion, we have demonstrated a few-hole DQD in an undoped (100) oriented GaAs/AlGaAs heterostructure.
The device shows good charge stability and negligible hysteresis with respect to gate voltage.
The interdot coupling can be tuned over a wide range, controling the transition from a large single dot to two well-isolated quantum dots.
Using charge sensing we show that the dot can be completely emptied of holes and operated in the few-hole regime.
The device may provide a means for future experiments focusing on manipulation of single hole spins in GaAs quantum dots.

\begin{acknowledgments}
We would like to thank J. Dominguez for assistance with sample fabrication.
This work was performed at the Center for Integrated Nanotechnologies, 
a U.S. DOE, Office of Basic Energy Sciences user facility, and Sandia National Laboratories, a multi-program laboratory operated by Sandia Corporation, 
a wholly owned subsidiary Lockheed-Martin Company, for the U. S. Department of Energy's National Nuclear Security Administration under Contract No. DE-AC04-94AL85000.
\end{acknowledgments}


\begin{thebibliography} {99}

\bibitem{Loss} D. Loss and D. P. DiVincenzo, Phys. Rev. A \textbf{57}, 120 (1998).
\bibitem{Petta} J. R. Petta, A. C. Johnson, J. M. Taylor, E. A. Laird, A. Yacoby, M. D. Lukin, C. M. Marcus, M. P. Hanson, A. C. Gossard,
Science \textbf{309}, 2180 (2005).
\bibitem{Koppens} F. H. L. Koppens, C. Buizert, K. J. Tielrooij, I. T. Vink, K. C. Nowack, T. Meunier, L. P. Kouwenhoven, and L. M. K. Vandersypen,
Nature \textbf{442}, 766 (2006).
\bibitem{HansonRMP} R. Hanson, L. P. Kouwenhoven, J. R. Petta, S. Tarucha, and L. M. K. Vandersypen, Rev. Mod. Phys. \textbf{79}, 1217 (2007).

\bibitem{LossDecoherence} J. Fischer, W. A. Coish, D. V. Bulaev, and D. Loss, Phys. Rev. B \textbf{78}, 155329 (2008).
\bibitem{Brunner} D. Brunner, B. D. Geradot, P. A. Dalgarno, G. W\"{u}st, K. Karrai, N. G. Stoltz, P. M. Petroff, R. J. Warburton, Science \textbf{325}, 70 (2009).
\bibitem{Yamamoto} K. De Greve, P. L. McMahon, D. Press, T. D. Ladd, D. Bisping, C. Schneider, M. Kamp, L. Worshech, S. H\"{o}fling, A. Forchel, and Y. Yamamoto,
Nat. Phys \textbf{7}, 872 (2011).
\bibitem{Gammon} A. Greilich, S. G. Carter, D. Kim, A. S. Bracker, and D. Gammon, Nat. Photon. \textbf{5}, 702 (2011).

\bibitem{EDSR} D. V. Bulaev, and D. Loss, Phys. Rev. Lett. \textbf{98}, 097202 (2007).
\bibitem{ESDRInSbNanowire} V. S. Pribiag, S Nadj-Perge, S. M. Frolov, J. W. G. van der Berg, I. van Weperen, S. R. Plissard, E. P. A. M. Bakkers,
and L. P. Kouwenhoven, Nat. Nanotech. \textbf{8}, 170 (2013).

\bibitem{SiGeWires} Y. Hu, F. Kuemmeth, C. M. Lieber, and C. M. Marcus, Nat. Nanotech. \textbf{7}, 47 (2012).

\bibitem{conditional} I. van Weperen, B. D. Armstrong, E. A. Laird, J. Medford, C. M. Marcus, M. P. Hanson, A. C. Gossard, 
Phys. Rev. Lett. \textbf{107}, 030506 (2011).
\bibitem{Sachrajda} L. Gaudreau, G. Granger, A. Kam, G. C. Aers, S. A. Studenikin, P. Zawadzki, M. Pioro-Ladrie\`{e}re, Z. R. Wasilewski,
and A. S. Sachrajda, Nat. Phys. \textbf{8}, 54 (2012).

\bibitem{EnsslinSET} B. Grbi\'{c}, R. Leturcq, K. Ensslin, D. Reuter, and A. D. Wieck, Appl. Phys. Lett \textbf{87}, 232108 (2005).
\bibitem{Grbic} B. Grbi\'{c} \textit{et. al}, AIP Conf. Proc. \textbf{893}, 777 (2007).
\bibitem{Burke} A. M. Burke, D. Waddington, D. Carrad, R. Lyttleton, H. H. Tan, P. J. Reece, O. Klochan, A. R. Hamilton, A. Rai, D. Reuter, A. D. Wieck, and A. P. Micolich,
Phys. Rev. B \textbf{86}, 165309 (2012).

\bibitem{Kane} B. E Kane, L. N. Pfeiffer, K. W. West, and C. K. Harnett, Appl. Phys. Lett. \textbf{63}, 2132 (1993).
\bibitem{Lilly} H. Noh, M. P. Lilly, D. C. Tsui, J. A. Simmons, E. H. Hwang, S. Das Sarma, L. N. Pfeiffer, and K. W. West, Phys. Rev. B \textbf{68}, 165308 (2003);
H. Noh, M. P. Lilly, D. C. Tsui, J. A. Simmons, L. N. Pfeiffer, and K. W. West, Phys. Rev. B \textbf{68}, 241308(R) (2003).
\bibitem{TML} T. M. Lu, D. R. Luhman, K. Lai, D. C. Tsui, L. N. Pfeiffer, and K. W. West, Appl. Phys. Lett. \textbf{90}, 112113 (2007).

\bibitem{HamiltonSET} O. Klochan, J. C. H. Chen, A. P. Micolich, A. R. Hamilton, K. Muraki, and Y. Hirayama, Appl. Phys. Lett. \textbf{96}, 092103 (2010).

\bibitem{Ciorga} M. Ciorga, A. S. Sachrajda, P. Hawrylak, C. Gould, P. Zwadzki, S. Jullian, Y. Feng, and Z. Wasilewski,
Phys. Rev. B \textbf{61}, R16315 (2000).

\bibitem{Willett} R. L. Willett, M. J. Manfra, L. N. Pfeiffer, and K. W. West, Appl. Phys. Lett. \textbf{91}, 033510 (2007).

\bibitem{KouwenhovenReview} L. P. Kouwenhoven, C. M. Marcus, P. L. McEuen, S. Tarucha, R. M. Westervelt, and N. S. Wingreen, 
in \textit{Mesoscopic Electron Transport}, Proceedings of the Advanced Study Institute, edited by L. L. Sohn, L. P. Kouwenhoven, 
and G. Sch\'"{o}n (Kluwer, Dordercht, 1997).

\bibitem{Winkler} R. Winkler, 2003, \textit{Spin-orbit Coupling Effects in Two-Dimensional Electron and Hole Systems} (Springer-Verlag, Berlin).

\bibitem{TMLcyc} T. M. Lu, Z. F. Li, D. C. Tsui, M. J. Manfra, L. N. Pfeiffer, and K. W. West, Appl. Phys. Lett. \textbf{92}, 012109 (2008).

\bibitem{shallowQD} W. Y. Mak, F. Sfigakis, K. Das Gupta, O. Klochan, H. E. Beere, I. Farrer, J. P. Griffiths, G. A. C. Jones, A. R. Hamilton, and D. A. Ritchie,
Appl. Phys. Lett. \textbf{102}, 103507 (2013).

\bibitem{Pepper} K. J. Thomas, J. T. Nicholls, M. Y. Simmons, M. Pepper, D. R. Mace, and D. A. Ritchie, Phys. Rev. Lett. \textbf{77}, 135 (1996).
\bibitem{Hamiltonp7} A. R. Hamilton, R. Danneau, O. Klochan, W. R. Clarke, A. P. Micholich, L. H. Ho, M. Y. Simmons, D. A. Ritchie,
M. Pepper, K. Muraki, and Y. Hirayama, J. Phys. Condens. Matter \textbf{20}, 164205 (2008).
\bibitem{Ensslinp7} Y. Komijani, M. Csontos, I. Shorubalko, T. Ihn, K. Ensslin, Y. Meir, D. Reuter, and A. D. Wieck, Europhys. Lett. \textbf{91}, 67010 (2010).



\end{thebibliography}
\end{document}